# Giant second harmonic generation by engineering of double plasmonic resonances at nanoscale

Ming-Liang Ren, Si-Yuan Liu, Ben-Li Wang, Bao-Qin Chen, Jiafang Li and Zhi-Yuan Li*

Laboratory of Optical Physics, Institute of Physics, Chinese Academy of Sciences, P.O. Box 603, Beijing 100190, China

*Email address: lizy@aphy.iphy.ac.cn

Abstract

We have investigated second harmonic generation (SHG) from Ag-coated $LiNbO_3$ (LN) core-shell nanocuboids and found that giant SHG can occur via deliberately designed double plasmonic resonances. By controlling the aspect ratio, we can tune fundamental wave (FW) and SHG signal to match the longitudinal and transverse plasmonic modes simultaneously, and achieve giant enhancement of SHG by $3\times10^5$ in comparison to a bare LN nanocuboid and by about one order of magnitude to the case adopting only single plasmonic resonance. The underlying key physics is that the double-resonance nanoparticle enables greatly enhanced trapping and harvesting of incident FW energy, efficient internal transfer of optical energy from FW to SHW, and much improved power to transport the SHG energy from the nanoparticle to the far-field region. The proposed double-resonance nanostructure can serve as an efficient subwavelength coherent light source through SHG and enable flexible engineering of light-matter interaction at nanoscale.

Surface plasmon resonance (SPR) induced optical field enhancement and localization have been of great interests and widely applied in many optical phenomena [1,2], such as surface enhanced Raman scattering, near-field microscope, enhanced fluorescence emission and nonlinear interaction, e.g. second harmonic generation (SHG). In this regard, the enhancement of SHG has been observed in metal nanoparticles [3], a sharp metal tip [4], planar metallic structures [5-7], and multifrequency gold nano-antenna [8]. Recently, SHG has been achieved efficiently in a plasmonic nanocavity which consists of a noncentrosymmetric medium sphere [9] and nanowire [10] enclosed in a metal shell. The fundamental wave (FW) at frequency $\omega$ was observe to get strongly confined in nanocavities when interacting with the SPR mode of a metal-coated nanoparticle or nanowire and then overlap with the nonlinear medium very well, leading to more efficient SHG at frequency $2\omega$ (with an enhancement factor over 50) than the bare core (without the Ag shell). However, only FW is resonant with the nanoparticle in these cases. The direct effects of SPR on the SHG signal are still unclear. A natural question arises: How if both FW and SHW are simultaneously on-resonant with two SPR modes of nanoparticles?

In this work we will address this problem by insightful design. The most important thing is to look for a nanoparticle with double SPR modes. For this purpose, we turn to nanoparticles of anisotropic geometry, among which nanorods have attracted great attention due to their dimension induced wavelength tunability and polarization sensitivity [11-13]. One can control the aspect ratio of nanorods easily and obtain double SPR modes simultaneously, which are associated with the longitudinal and transverse SPR modes [14,15]. Moreover, the optical cross-sections of nanorods are much higher than those of nanospheres, and this has attributed nanorods promising features in wide applications like low-threshold surface plasmon amplification [16,17] and ultrafast optical devices [18]. The two SPR modes can be designed to enable their wavelengths matched to the absorption bands and emission bands of fluorescent molecules absorbed in gold nanorods [19] for enhancing their fluorescence intensity. It would be expected that this double-resonance mechanism should also play a very good role in enhancing SHG intensity of nanostructured materials.

To illustrate and confirm this hypothesis, we propose and design an anisotropic nanoparticle, an Ag-coated $LiNbO_3$ (LN) core-shell nanocuboid [Ag-LN in Fig. 1(a)] to further enhance nanoscale SHG, where longitudinal and transverse SPR modes are obtained simultaneously and match exactly the wavelengths of FW and SHG signal. Note that SHG occurs only from LN materials but not from Ag materials, but the SPR induced by Ag materials will significantly enhance SHG from LN the materials. We will show via numerical calculations that the SHG signal from Ag-LN with double SPR modes is much more efficient than the bare LN core ($>3\times10^5$X) where no SPR mode occurs [Fig. 1(b)].

Note that several theoretical methods such as the perturbation theory [20], nonlinear Mie theory [21] and reciprocity theory combined with linear Mie scattering [22], have been adopted to describe SHG cases in nanoparticles. However, most of these methods are limited to handle spherical nanoparticles, and a universal model is still rare for nanoparticles in arbitrary shapes. Here, we develop a nonlinear discrete-dipole approximation (DDA) method to analyze SHG case conveniently in arbitrary nanoparticles in the framework of rigorous Maxwell's equations solution. As is known, the DDA method is an effective and powerful tool to calculate the linear optical properties of nanoparticles, such as extinction, scattering and absorption [23, 24]. This approach is efficient equally for nonlinear optical problem solutions.

Assume a plane wave FW at the frequency of $\omega_1$ is incident upon a nonlinear particle. The FW electric field $\mathbf{E}_1$ satisfies Maxwell's equation

$$\nabla\times\nabla\times\mathbf{E}_1 - k_1^2\mathbf{E}_1 = k_1^2\mathbf{P}_1, \tag{1}$$

where $\mathbf{P}_1$ is the linear polarization of FW, $k_1 = \omega_1/c$, $c$ is light speed in vacuum. To use the DDA [23, 24] to solve Eq. (1), the particle is assumed as a cubic array of $N$ electric dipoles, and the total electric field in the $j^{th}$ dipole can be described by

$$\mathbf{E}_{1,j} = \mathbf{E}_{10,j} - \sum_{k\neq j}^{N}\mathbf{A}_{jk}\mathbf{P}_{1,k}, \tag{2}$$

where $\mathbf{E}_{10,j} = \mathbf{E}_{10}\exp(ik_1\mathbf{n}_j\cdot\mathbf{r}_j)$, $\mathbf{P}_{1,j} = \alpha_{1,j}\mathbf{E}_{1,j}$, $\mathbf{n}_j = \mathbf{r}_j/|\mathbf{r}_j|$, $\alpha_{1,j}$ is the polarizability of the $j^{th}$ dipole at $\omega_1$. $-\mathbf{A}_{jk}\mathbf{P}_{1,k}$ is the secondary radiation electric field at $\mathbf{r}_j$ from dipole $\mathbf{P}_{1,k}$ at $\mathbf{r}_k$, and it is given as,

$$\mathbf{A}_{jk}\mathbf{P}_{1,k} = \frac{\exp(ik_1 r_{jk})}{r_{jk}^3}\{k_1^2\mathbf{r}_{jk}\times(\mathbf{r}_{jk}\times\mathbf{P}_{1,k}) + \frac{1-ik_1 r_{jk}}{r_{jk}^2}[r_{jk}^2\mathbf{P}_{1,k} - 3\mathbf{r}_{jk}(\mathbf{r}_{jk}\cdot\mathbf{P}_{1,k})]\}, \tag{3}$$

where $\mathbf{r}_{jk} = \mathbf{r}_j - \mathbf{r}_k$ and $r_{jk} = |\mathbf{r}_{jk}|$. Each element $\mathbf{A}_{jk}$ is a 3×3 matrix. Defining

$\mathbf{A}_{jj} = \alpha_j^{-1}$, Eq. (2) can be written into a simple algebraic form of

$$\mathbf{E}_{10,j} = \sum_{k=1}^{N} \mathbf{A}_{jk} \mathbf{P}_{1,k}. \qquad (4)$$

Once all the dipoles $\mathbf{P}_{1,k}$ are calculated self-consistently by solving the linear equations Eq. (4), the total scattering cross-section can be given by a spherical integral as

$$C_{sca} = \frac{k_1^4}{|\mathbf{E}_{10}|^2} \int d\Omega \left| \sum_{k=1}^{N} [\mathbf{P}_{1,k} - \mathbf{n}_k (\mathbf{n}_k \cdot \mathbf{P}_{1,k})] \exp(-ik_1 \mathbf{n}_k \cdot \mathbf{r}_k) \right|^2, \qquad (5)$$

and the extinction cross-section is

$$C_{ext} = \frac{4\pi k_1}{|\mathbf{E}_{10}|^2} \sum_{k=1}^{N} \mathrm{Im}(\mathbf{E}_{10,k}^* \cdot \mathbf{P}_{1,k}). \qquad (6)$$

In the SHG case, the SHG signal at the frequency of $\omega_2 = 2\omega_1$ has its electric field $\mathbf{E}_2$ satisfying the following nonlinear coupled equation [25, 26],

$$\nabla \times \nabla \times \mathbf{E}_2 - k_2^2 \mathbf{E}_2 = k_2^2 \mathbf{P}_2 + k_2^2 \mathbf{P}^{(2)}, \qquad (7)$$

where $k_2 = \omega_2/c$, $\mathbf{P}_2$ is the linear polarization of the SHG signal, and $\mathbf{P}^{(2)}$ denotes the second-order nonlinear polarization. Similar to the linear case in Eq. (2), the total electric field of the SHG signal in the $j^{\mathrm{th}}$ dipole is derived as

$$\mathbf{E}_{2,j} = \mathbf{E}_{20,j} - \sum_{k \neq j}^{N} \mathbf{B}_{jk} \mathbf{P}_{2,k}, \qquad (8)$$

Assume $\alpha_{2,j}$ is the polarizability of the $j^{\mathrm{th}}$ dipole at $\omega_2$ and $\mathbf{P}_{2,j} = \alpha_{2,j} \mathbf{E}_{2,j}$. Similar to $\mathbf{A}_{jk}$, $\mathbf{B}_{jk}$ reflects the interaction between the $j^{\mathrm{th}}$ and $k^{\mathrm{th}}$ dipoles at $\omega_2$ and follows,

$$\mathbf{B}_{jk} \mathbf{P}_{2,k} = \frac{\exp(ik_2 r_{jk})}{r_{jk}^3} \{k_2^2 \mathbf{r}_{jk} \times (\mathbf{r}_{jk} \times \mathbf{P}_{2,k}) + \frac{1 - ik_2 r_{jk}}{r_{jk}^2} [r_{jk}^2 \mathbf{P}_{2,k} - 3\mathbf{r}_{jk}(\mathbf{r}_{jk} \cdot \mathbf{P}_{2,k})]\}, \qquad (9)$$

In Eq. (8), $\mathbf{E}_{20,j}$ is the electric filed that is due to $\mathbf{P}^{(2)}$. It can also be written in the context of DDA as

$$\mathbf{E}_{20,j} = \mathbf{P}_j^{(2)}/\alpha_{2,j} - \sum_{k \neq j}^{N} \mathbf{B}_{jk} \mathbf{P}_k^{(2)}, \tag{10}$$

where $\mathbf{P}_j^{(2)}/\alpha_{2,j}$ represents the self-induced radiation field by $\mathbf{P}_j^{(2)}$ at $\mathbf{r}_j$. $-\mathbf{B}_{jk}\mathbf{P}_k^{(2)}$ has a similar expression as Eq. (9) where $\mathbf{P}_{2,k}$ is replaced by $\mathbf{P}_k^{(2)}$, and it represents the secondary radiation electric field induced by nonlinear dipole $\mathbf{P}_k^{(2)}$ at $\mathbf{r}_k$. Each nonlinear dipole $\mathbf{P}_j^{(2)}$ can be given as

$$\mathbf{P}_j^{(2)} = \frac{d^3}{4\pi\varepsilon_3}(\frac{3\varepsilon_3}{\varepsilon_1 + 2\varepsilon_3})^2 \chi^{(2)} : \mathbf{E}_{1,j}\mathbf{E}_{1,j}, \tag{11}$$

where $\varepsilon_1$ stands for the dielectric constant of studied medium, while $\varepsilon_3$ for the background medium. Note that $\mathbf{E}_{1,j}$ specifies the external filed of the $j^{th}$ dipole, thus the factor $[3\varepsilon_3/(\varepsilon_1 + 2\varepsilon_3)]^2$ should be considered in Eq. (11).

Now we have made relevant deviations of the nonlinear DDA method. Note that the conversion efficiency of SHG is usually low in nanoparticles, the undepleted pump approximation (UPA) is adopted. The nonlinear DDA method can be summarized as follows: when FW is incident upon the particle, each dipole $\mathbf{P}_{1,j}$ and the total electric field $\mathbf{E}_{1,j}$ are calculated by means of DDA [shown in Eqs. (1)-(6)]; then the nonlinear dipole $\mathbf{P}_j^{(2)}$ is induced by $\mathbf{E}_{1,j}$ in the second-order nonlinear process, and regarded as the radiation source of the SHG signal. Subsequently, $\mathbf{E}_{20,j}$ and $\mathbf{P}_{2,j}$ are solved self-consistently [shown in Eqs. (7)-(11)]. Once the total dipole distribution of SHW $\mathbf{P}$ $(= \mathbf{P}_{2,j} + \mathbf{P}_j^{(2)})$ is obtained, we can give the scattering SHG power by

$$W_{sca} \propto k_2^4 \int d\Omega \left| \sum_{k=1}^{N} [\mathbf{P}_k - \mathbf{n}_k(\mathbf{n}_k \cdot \mathbf{P}_k)]\exp(-ik_2\mathbf{n}_k \cdot \mathbf{r}_k) \right|^2. \tag{12}$$

This formula is analogous to Eq. (5) and is used to evaluate the SHG radiation generated from the particle.

Now we take a closer look at Eq. (11), which is one of the key ingredients of the nonlinear DDA method. It describes the nonlinear polarization ($\mathbf{P}^{(2)}$) which irradiates the SHG signal. $\mathbf{P}^{(2)}$ depends both on the local FW electric field and on the nonlinear susceptibility by the following matrix equation [26],

$$\begin{pmatrix} P_x^{(2)} \\ P_y^{(2)} \\ P_z^{(2)} \end{pmatrix} = \begin{pmatrix} d_{11} & d_{12} & d_{13} & d_{14} & d_{15} & d_{16} \\ d_{21} & d_{22} & d_{23} & d_{24} & d_{25} & d_{26} \\ d_{31} & d_{32} & d_{33} & d_{34} & d_{35} & d_{36} \end{pmatrix} \begin{pmatrix} E_{1x}^2 \\ E_{1y}^2 \\ E_{1z}^2 \\ 2E_{1y}E_{1z} \\ 2E_{1z}E_{1x} \\ 2E_{1x}E_{1y} \end{pmatrix}, \qquad (13)$$

where $\{d_{ij}\}$ depicts the second-order nonlinear tensor of the nanoparticle material. In the current study, the nonlinear material is the LN core, where only $d_{15}$, $d_{16}$, $d_{21}$, $d_{22}$, $d_{24}$, $d_{31}$, $d_{32}$, and $d_{33}$ are non-vanishing. As shown in Eqs. (7)-(13), the SHG process can be described in two steps: the nonlinear polarization generation in the LN core (depending on FW and $\{d_{ij}\}$) and SHG radiation out of the Ag shell. Simply speaking, the SHG radiation power of the core-shell nonlinear nanoparticle relies on two apparent factors: the efficiency to create a strong local nonlinear polarization and the efficiency to transport the local SHG energy from the near-field region of nanoparticle to its far-field region. In the first factor, the local electric field and the nonlinear susceptibility both play important roles. SPR resonant with the incident FW will induce a greatly enhanced linear local field and contribute much to the first factor, while SPR resonant with the SHW will contribute much to the second factor. A giant enhancement of SHG should be expected when a nonlinear nanoparticle possesses double SPRs that match both with the FW and SHW simultaneously. To achieve this highly desirable goal, deliberate design on the geometric configuration must be implemented.

We assume the pump amplitude (of FW) is 1.0 V/m and the second-order nonlinear coefficient is $d_0$=4 pm/V. As shown in Fig. 1(a), the plane-wave FW propagates in the $x$-axis and is polarized in the $yz$ plane. We only consider SHG from the LN core of which the second-order nonlinear coefficient is much larger than the metal shell and surface, as is the case for Ag shell, and suppose this core-shell nanoparticle is embedded in the water background ($\varepsilon_3$=1.33). The refractive index of LN is chosen as 2.17 near 900 nm, while 2.23 near 450 nm. By means of the nonlinear DDA method developed above, we have calculated and compared SHG in three different cases [Fig. 1(b)], (1) Ag-LN where $y$-polarized FW ($E_{1y}$) excites $z$-polarized SHG signal ($E_{1z}$) by utilizing the nonlinear coefficient of $d_{32}$ ($E_{1y}E_{1y}$-$E_{2z}$, or $yy$-$z$), (2) Ag-LN ($E_{1y}E_{1y}$-$E_{1y}$, or $yy$-$y$ with $d_{22}$) and (3) bare LN core ($yy$-$y$). It is seen that the SHG signal is very poor from the bare LN and has no any feature. However, the nonlinear signal exhibits an obvious peak (~450 nm) in Ag-LN with *yy-z*, much stronger (>3×10$^5$X) than the bare LN. More interestingly, the *yy-z* process seems to be more efficient (~10X) than *yy-y* even if $d_{32}=d_{22}=d_0$. Here we only consider $d_{32}$ and $d_{22}$ for comparison.

In order to explore the underlying mechanism, we have calculated the extinction cross-section ($C_{ext}$) of these structures (Ag-LN and bare) in Fig. 2. Under the *y*-polarized excitation (along long side), there is a resonance mode around 900 nm in Ag-LN, corresponding to the longitudinal SPR mode (LSPR) [Fig. 2(a)], while no remarkable feature is observed in the bare LN [Figs. 2(c) and 2(d)]. Therefore if FW is tuned to match LSPR exactly (~900 nm) in Ag-LN, it can get intensified greatly in the *yy-y* process and then enhance the nonlinear polarization generation, which serves as an efficient light source of SHG radiation [>10$^4$X enhancement achieved in comparison to the bare LN in Fig. 1(b)]. Note that there is no peak of $C_{ext}$ around 450 nm under the *y*-polarized excitation [Inset of Fig. 2(a)]. Only FW is on-resonance in the *yy-y* process (single SPR mode) while the SHG signal is off-resonance, similar to previous study [9, 10]. Figure 2(b) shows two resonance modes take place in the short wavelength region if the excitation light is polarized along the *z*-axis (the short side). One of them is located near 450 nm (double frequency of FW at 900 nm) and is related to the transverse SPR mode (TSPR), which could also facilitate the light emission [27]. In the *yy-z* process, FW is tuned at the LSPR mode (similar to *yy-y*) and the generated SHG signal matches the TSPR mode exactly. Therefore the double SPR modes in this case (*yy-z*) result in much stronger SHG radiation (~10X) than the single SPR mode (*yy-y*) [Fig. 1(b)].

To further understand these differences, we also show the FW electric field intensity distribution ($I_\omega$) of Ag-LN and bare LN in different cases [Fig. 3]. At the LSPR mode (~900 nm) of Ag-LN, the light is strongly confined in the plasmonic nanocavity, ~250X amplification in the most LN body and >500X in the Ag/LN interface [Fig. 3(a)], which is very constructive to the SHG process (strong field overlapping with the nonlinear core). In contrast, the light concentration is very poor (<5X) in the bare LN under the same excitation condition [Fig. 3(c)], corresponding to the inefficient SHG [Fig. 1(b)]. Moreover, the TSPR mode (~450 nm) in Ag-LN exhibits much better concentration of light in the nanocavity (>10X) [Fig. 3(b)] than the bare LN under the *z*-polarized excitation [Fig. 4(d)]. Therefore, Ag-LN is a better candidate for efficient SHG (due to the double SPR modes) than the bare LN. In this system with double SPR modes, the overlap between FW and SH field modal profile (e.g. *yy-z*), $\kappa = \left| \int d_{32} E_{1y} E_{1y} E_{2z} dV \right|$, is another critical parameter to affect the SHG radiation [7] and seems to be good in our case [Figs. 3(a) and 3(b)]. Note that the SHG signal is generated only from the LN core, the field distribution is only plotted in the LN core (even in Ag-LN).

To study the anisotropic properties of SHG as naturally expected for the geometrically anisotropic Ag-LN core-shell nanoparticle, we tune the FW excitation polarization angle from $\theta=0°$ (along the *y*-axis) to 90° (along the *z*-axis) and show the calculated *z*-polarized SHG signal from Ag-LN by setting FW at 900 nm in Fig. 4(a). When $\theta=0°$, FW is polarized along the *y*-axis and then matches the LSPR mode [Fig. 2(a)], leading to the efficient SHG radiation [Fig. 1(b)]. However, there is no resonance mode near 900 nm under the *z*-polarized excitation ($\theta=90°$) [Inset of Fig. 2(b)]. Therefore, the SHG signal is observed to decrease monotonically as the polarization angle increases [Fig. 4(a)]. Here the nonlinear coefficients $d_{32}$ (=$d_0$) and $d_{33}$ (=$7d_0$) are involved. If FW is polarized along the *x*-axis (isotropic or identical to the *z*-axis due to the same side length) and propagates along the *y*-axis, the SHG signal (*xx-y* and *xx-z*) is very poor in comparison to the *yy-z* process [Fig. 1(b)] but still exhibits some peak feature [Fig. 4(b)], very similar to the FW behavior in the extinction spectrum [Inset of Fig. 2(b)] where a small peak is observed (weakly-resonant). Furthermore, the *xx-z* process ($d_{31}$) appears to be more efficient than *xx-y* ($d_{21}$) even if $d_{31}=d_{21}=d_0$ as a result of the TSPR mode excited by the *z*-polarized SHG signal [Fig. 2(b)]. Another interesting thing is that the SHG signal in the *xx-z* process is maximum at ~445 nm, red-shifting from the *xx-y* case (~440 nm) but blue-shifting from the TSPR mode (~450 nm), which is probably due to the joint contribution of the weakly-resonant FW and TSPR mode in the *xx-z* process.

In the above discussion we tune FW at the longitudinal SPR mode (~ 900 nm) and the SHG signal at the transverse one (~450 nm), and achieve nanoscale SHG efficiently in the silver-coated nonlinear nanocuboid. Note that the nonlinear coefficients utilized are assumed to be the same for comparison. One may achieve even higher conversion efficiency of SHG by combining double SPR modes with the largest nonlinear coefficient (e.g. $d_{33}$ in LN) if adjusting the orientation of nanoparticles. Moreover, we only consider SHG in an isolated nanocuboid here for simplification, which holds true in the system with sparse nanocuboids. However if nanocuboids are packed densely and aligned by some methods, such as a stretched-film method [14,15], plasmonic coupling or interaction between nanocuboids could be strong and remarkable [27], probably contributing more to the SHG radiation. As double SPR modes closely rely on the aspect ratio of a nanocuboid [11-19,28], one can tune them to desired wavelengths for efficient SHG and enhanced fluorescence radiation [19]. If the nanocuboid is anisotropic in *x*-, *y*- and *z*-axis, three modes could be obtained simultaneously and therefore be adopted to enhance third harmonic generation (THG). Another point that is worthwhile to mention is that in the nanocuboid particle, the TSPR mode is much weaker than the LSPR mode in the strength of both optical cross section and field enhancement. It is expected that by considering other double SPR nanosystems with equally strong resonant strength, the SHG can become even stronger by orders of magnitude.

In conclusion, a silver-coated nonlinear nanocuboid has been proposed and demonstrated as an efficient subwavelength coherent light source through SHG, which can help to study various optical phenomena in nanoscale size. The study is based on numerical simulation using a home-made nonlinear DDA method that can rigorously solve SHG from arbitrary nonlinear nanoparticles. We have found that when FW and SHW are tuned to match the longitudinal and transverse SPR modes of the nanocuboid particle simultaneously, nanoscale SHG radiation can be enhanced by ~10X in comparison to the single SPR mode (*yy-y*) and >3×10$^5$X to the bare LN [Fig. 1(b)]. In general, double SPR modes in anisotropic nanocuboids can be harnessed to achieve efficient nanoscale SHG, strong luminescence, enhanced fluorescence absorption and emission, and other applications like high-resolution cell imaging and biomedical labeling [29], and will open a new avenue to manipulate light-matter interaction in nanoscale plasmonic systems [31-33].

This work was supported by the State Key Development Program for Basic Research of China at No. 2013CB632704, and the National Natural Science Foundation of China at Nos. 11104342 and 11374357.

Figure captions

**Fig. 1** (Color online) (a) Schematic diagram of a silver-coated nanocuboid. The nonlinear core, $16\times72\times16$ nm$^3$, is lithium niobate (LiNbO$_3$, LN) which is coated by a 5 nm thick silver shell. The fundamental wave (FW, $\omega$) propagates along the *x*-direction, and excites second harmonic generation (SHG, $2\omega$) single in all directions. (b) Calculated SHG signal in three different cases, Ag-coated LiNbO$_3$ nanocuboid (Ag-LN, *yy-z* utilizing $d_{32}$), Ag-LN (*yy-y*, $d_{22}$), and bare LN (*yy-y*, $d_{22}$) for comparison. Here *yy-z* representss that the SHG signal with *z*-component ($E_{2z}$) is excited by FW with $E_{1y}$. Assume $d_{32}=d_{22}=d_0$. $P_0$ is the unit power.

**Fig. 2** (Color online) Extinction cross-section ($C_{ext}$) of Ag-coated and bare LN nanocuboid. (a) Ag-LN under the *y*-polarized excitation. Insets: diagram of the excitation polarization and enlarged figure of $C_{ext}$ wihtin the wavelength range of 400-500 nm. (b) Ag-LN under the *z*-polarized excitation. Similar to Panel (a), one of insets shows the enlarged figure of $C_{ext}$ within 800-1000 nm. Bare LN under the *y*-polarized (c) and *z*-polarized (d) excitation.

**Fig. 3** (Color online) Field distribution (*yz*-plane) of Ag-coated and bare LN nanocuboid at the wavelengths of 900 nm and 450 nm. (a) Ag-LN under the *y*-polarized excitation and at 900 nm (resonance). (b) Ag-LN under the *z*-polarized excitation and at 450 nm (resonance). (c) Bare LN under the *y*-polarized excitation and at 900 nm. (d) Bare LN under the *z*-polarized excitation and at 450 nm. Here only the LN core (of Ag-LN) is plotted.

**Fig. 4** (Color online) (a) Excitation polarization dependence of the *z*-polarized SHG signals. Inset: polarization angle of FW (*yz*-plane), $\theta$, is set with respect to the *y*-axis. (b) Calculated SHG signal in the cases of *xx-z* (utilizing $d_{31}$) and *xx-y* ($d_{21}$) when FW is polarized in the *xz*-plane (propagating along the *y*-axis). Assume $d_{31}= d_{21}=d_0$.

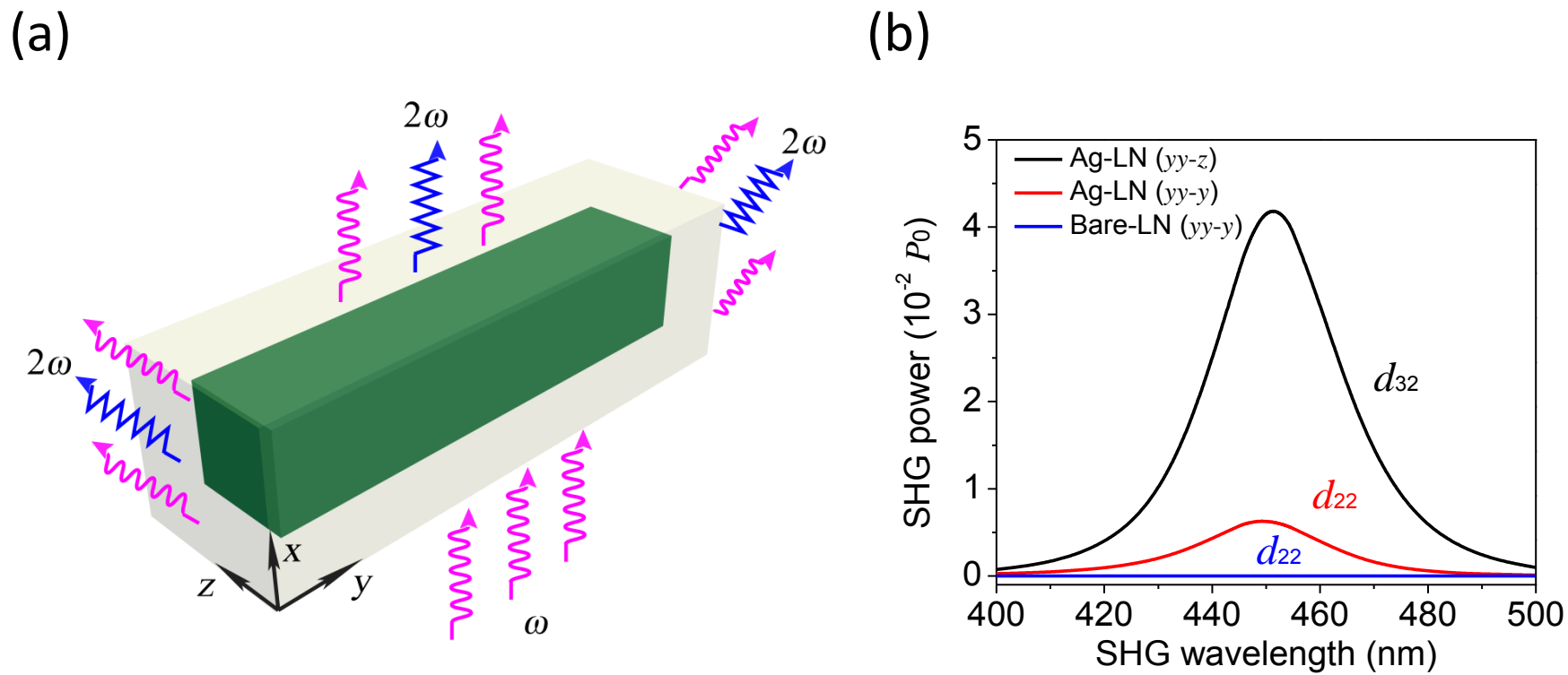

Fig. 1

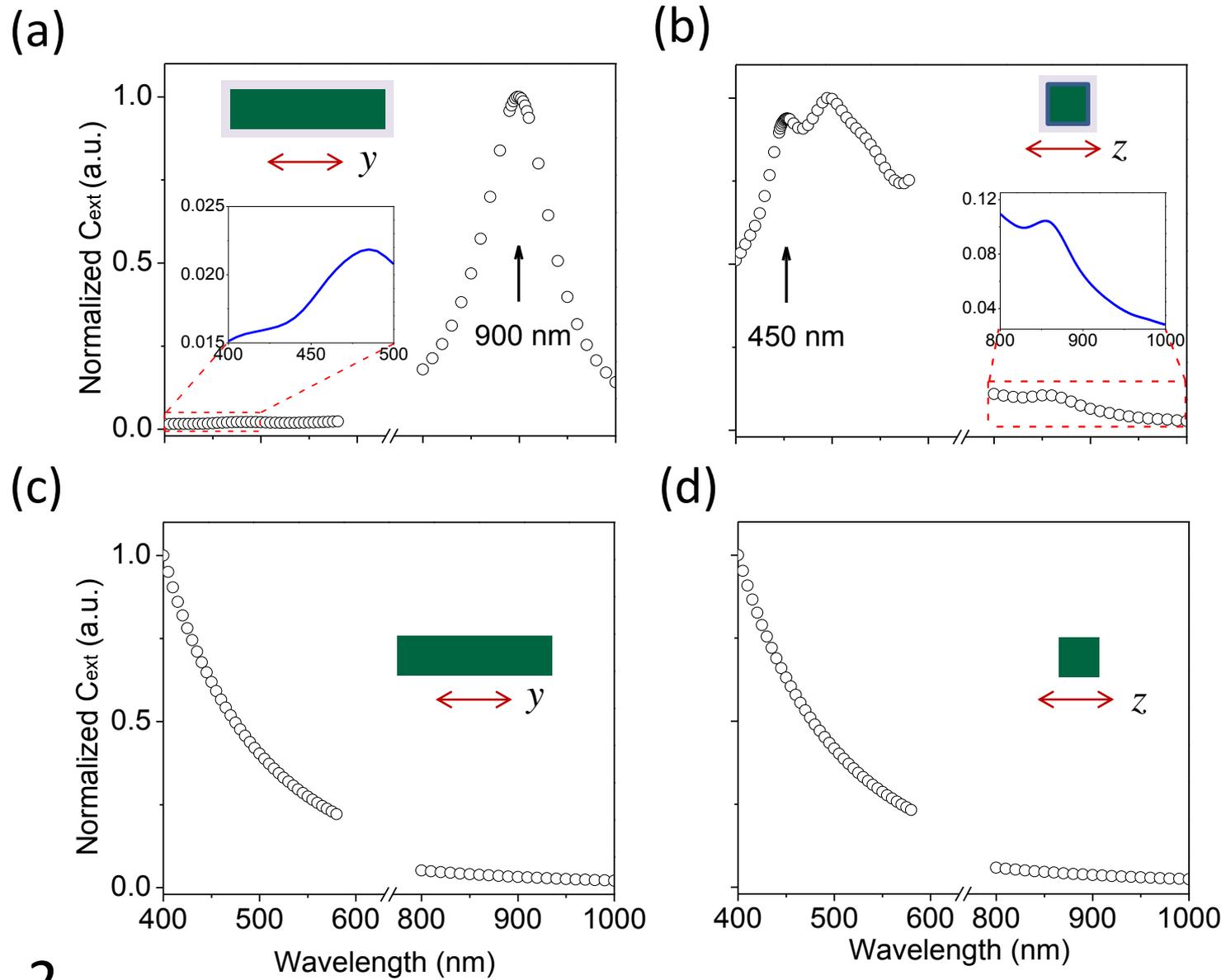

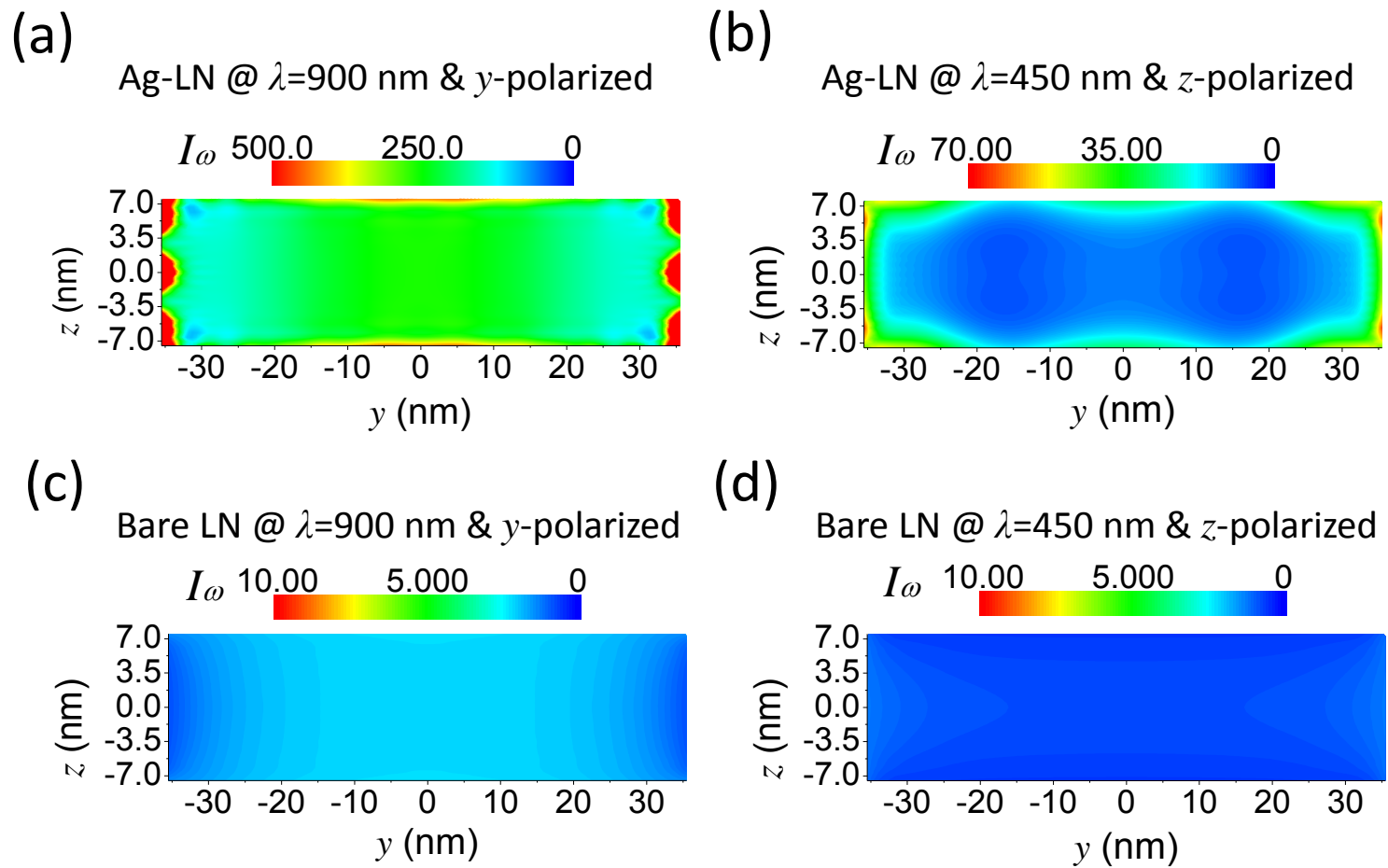

Fig. 3

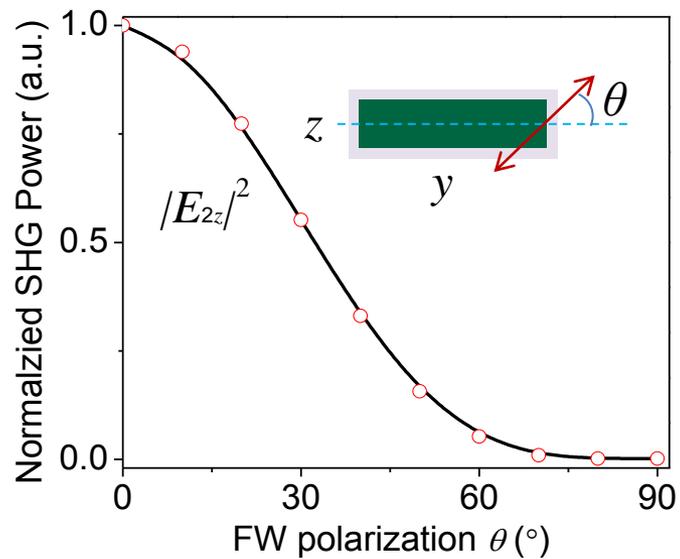 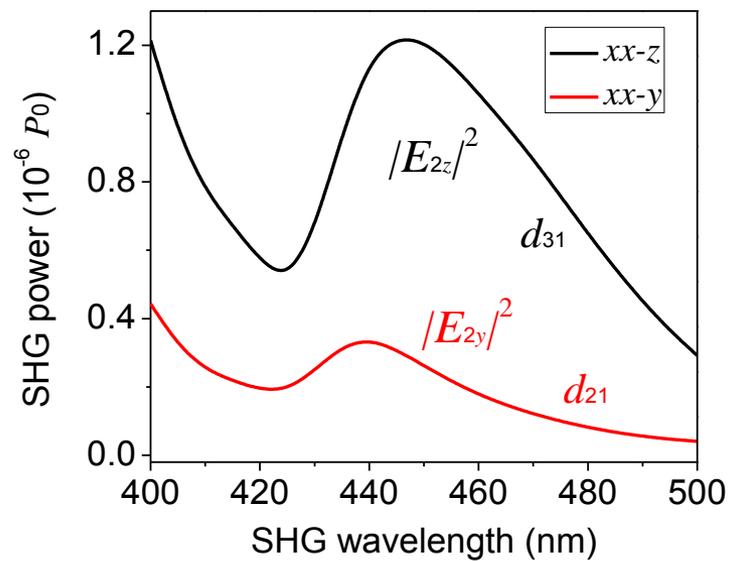

Fig. 4